\documentclass{bmcart}

\usepackage{amsthm,amsmath}

\usepackage[backend=biber,style=alphabetic]{biblatex}
\IfFileExists{\jobname.aux}{%
	\RequirePackage{hyperref}\usepackage{orcidlink}\hypersetup{colorlinks = false,}
}{%
}

\usepackage[utf8]{inputenc} 

\startlocaldefs

\usepackage{comment}
\usepackage{makecell}
\usepackage{placeins}
\usepackage[figuresright]{rotating}
\usepackage{subcaption}
\usepackage[nointegrals]{wasysym}

\usepackage{graphicx}%
\usepackage{multirow}%
\usepackage{amsmath,amssymb,amsfonts}%
\usepackage{amsthm}%
\usepackage{mathrsfs}%
\usepackage[title]{appendix}%
\usepackage{xcolor}%
\usepackage{textcomp}%
\usepackage{manyfoot}%
\usepackage{booktabs}%
\usepackage{algorithm}%
\usepackage{algorithmicx}%
\usepackage{algpseudocode}%
\usepackage{listings}%
\usepackage{tikz}

\theoremstyle{thmstyleone}%
\theoremstyle{thmstyletwo}%

\theoremstyle{thmstylethree}%

\newcommand{\beckOnline}{\emph{beck-online}}
\newcommand{\challengeAccessibility}{Accessibility Issues}
\newcommand{\challengeAI}{AI Usage}
\newcommand{\challengeComplexity}{Navigational Complexity}
\newcommand{\challengeInefficiency}{Inefficiency of Current Interfaces}
\newcommand{\challengeReliance}{Reliance on Tacit Knowledge}
\newcommand{\juris}{\emph{juris}}

\newcommand{\tasksFundamental}{$T_1$}
\newcommand{\tasksRelationships}{$T_2$}
\newcommand{\tasksReasoning}{$T_3$}
\newcommand{\tasksSpecialized}{$T_S$}
\newcommand{\workflowFirstPhase}{$W_\texttt{P1}$}
\newcommand{\workflowSecondPhase}{$W_\texttt{P2}$}
\newcommand{\workflowThirdPhase}{$W_\texttt{P3}$}

\IfFileExists{\jobname.aux}{%

}{%
	\newcommand{\orcidlink}{{}}
	\let\autoref\ref
	\newcommand{\href}{{}}
	\newcommand{\hyperref}{{}}
}

\endlocaldefs

\begin{document}

\begin{frontmatter}

\begin{fmbox}
\dochead{Research Note}

\title{Challenges and Opportunities for Visual Analytics in Jurisprudence}

\author[
addressref={aff1},                   
corref={aff1},                       
email={daniel.fuerst@uni-konstanz.de}   
]{\inits{D.F.}\fnm{Daniel} \snm{Fürst\,\orcidlink{0000-0002-0407-2867}}}
\author[
addressref={aff2},
email={elassady@ethz.ch}
]{\inits{M.E.}\fnm{ Mennatallah} \snm{El-Assady\,\orcidlink{0000-0001-8526-2613}}}
\author[
addressref={aff1},
email={keim@uni-konstanz.de}
]{\inits{D.A.K}\fnm{Daniel A.} \snm{Keim\,\orcidlink{0000-0001-7966-9740}}}
\author[
addressref={aff1},
email={max.fischer@uni-konstanz.de}
]{\inits{M.T.F.}\fnm{Maximilian T.} \snm{Fischer\,\orcidlink{0000-0001-8076-1376}}}


\address[id=aff1]{
  \orgdiv{Department of Computer and Information Science},             
  \orgname{University of Konstanz},          
  \city{Konstanz},                              
  \cny{Germany}                                    
}
\address[id=aff2]{
  \orgdiv{Department of Computer Science},             
  \orgname{ETH Zurich},          
  \city{Zurich},                              
  \cny{Switzerland}                                    
}


\end{fmbox}

\begin{abstractbox}

\begin{abstract} 
Legal exploration, analysis, and interpretation remain complex and demanding tasks, even for experienced legal scholars, due to the domain-specific language, tacit legal concepts, and intentional ambiguities embedded in legal texts.
In related, text-based domains, Visual Analytics~(VA) has become an indispensable tool for navigating documents, representing knowledge, and supporting analytical reasoning.
However, legal scholarship presents distinct challenges: it requires managing formal legal structure, drawing on tacit domain knowledge, and documenting intricate and accurate reasoning processes~--~needs that current VA system designs for law fail to address adequately. 
We identify and describe key challenges and underexplored opportunities in applying VA to law, exploring how these technologies might better serve the legal domain.
Interviews with nine legal experts reveal that current legal information retrieval interfaces do not adequately support the navigational complexity of law, often forcing users to rely on internalized legal expertise instead.
To address this gap, we identify a three-phase workflow for legal experts, which highlights opportunities for VA to support legal reasoning through knowledge externalization and provenance tracking, leveraging tree-, graph-, and hierarchy-based visualizations.
Through this contribution, our work establishes a user-centered VA workflow for the legal domain, recognizing tacit legal knowledge as a critical element of sense-making and insight generation, and situates these contributions within a broader research agenda for VA in law and other text-based disciplines.
\end{abstract}



\end{abstractbox}
%

\end{frontmatter}
\begin{tikzpicture}[remember picture,overlay]
\node[text width=14cm, align=justify, anchor=north,yshift=-14pt, execute at begin node=\setlength{\baselineskip}{1em}] at (current page.north)
    {\scriptsize{© The Authors 2025. This is the author's version of the article that has been published in Springer's Artificial Intelligence and Law journal. The final version of this article should be cited as: \textit{Daniel Fürst, Mennatallah El-Assady, Daniel A. Keim, Maximilian T. Fischer}: Challenges and Opportunities for Visual Analytics in Jurisprudence, Artificial Intelligence and Law, Springer. doi: \href{https://dx.doi.org/10.1007/s10506-025-09494-2}{10.1007/s10506-025-09494-2} (to appear).}};
\end{tikzpicture}

\section{Introduction}
\label{section:foundations-and-background}

Legal text data is vast~\cite{fobbeCorpusDeutschenBundesrechts2024, fobbeCorpusEntscheidungenBundesverfassungsgerichts2024, fobbeCorpusDrucksachenDeutschen2021} but inherently structured~\cite{ruethersRechtstheorieUndJuristische2022}. It combines \emph{domain-specific language}~\cite{ruethersRechtstheorieUndJuristische2022} and \emph{precise phraseology} with intentional \emph{ambiguity} at times~\cite{ruethersRechtstheorieUndJuristische2022}. This ambiguity allows for statutory interpretation~\cite{ruethersRechtstheorieUndJuristische2022}, enabling laws to evolve alongside societal values.
For instance, in 2018, the German Federal Constitutional Court interpreted Art. 2 Abs. 2 Satz 1 GG  to include protection against the dangers of climate change~\cite{bundesverfassungsgerichtBundesverfassungsgerichtDecisionsConstitutional2021}, where the legal text merely states that ``...every person shall have the right to life and physical integrity...'' This interpretation underscores that considering legal norms in isolation is insufficient; statutory interpretation also relies on an extensive understanding of fundamental and specific legal principles~\cite{ruethersRechtstheorieUndJuristische2022}.

These legal principles are formed and conceived within individual jurisdictions and legal systems.
Several legal systems co-exist, with common law and civil law being predominant~\cite{glennComparativeLegalFamilies2006}, while the latter is the most widespread.
The Anglo-American common law system relies on judicial precedents and case law, whereas the civil law system, rooted in codes and statutes, lends itself to more structured analysis.
These differences are the focus of comparative law studies, reaching back centuries~\cite{Schmitthoff.ComparativeLaw.1939, Zimmermann.ComparativeLawHistory.1996}.
Seminal work~\cite{Fikentscher.MethodenRechtsvergleich.1975, David.MajorLegalSystems.1978} has analyzed comparability of methodologies and approaches, finding that legal practices can differ between jurisdictions and even within legal systems.
In particular, with tighter integrations in, for example, the European Union, contrasting is crucial~\cite{Schadbach.BenefitsComparativeLaw.1998}.
While differences like inductive vs. deductive reasoning, codification vs. precedent referencing, and systematic codes vs. narrative case opinions affect workflows, recent research has shown that broader concepts like goals, methods, and structures remain similar across countries within one legal system~\cite{Rossi.Rechtsvergleichung.2017, Reimann.HandbookComparativeLaw.2019, Zweigert.Rechtsvergleichung.2024}.

To navigate the landscape of the legal domain, scholars have developed working methods to aid in the interpretation of legal materials, including implicit structures, schemata, and principles~\cite[p.~450]{ruethersRechtstheorieUndJuristische2022}, which are trained and ingrained through years of demanding studies.
Visualizations can enhance these methods by improving comprehension and communicating mental models~\cite{passalacquaUsingVisualTechniques1997, brunschwigVisualLawVisual2014}.
These approaches provide a foundational starting point for examining, with the help of domain experts in law, how Visual Analytics~(VA)~\cite{keimVisualAnalyticsDefinition2008} can enhance legal research practices.
While the legal domain usually employs visualization for a specific purpose on selected occasions~\cite{passalacquaUsingVisualTechniques1997, brunschwigVisualLawVisual2014}, in Human-Computer-Interaction~(HCI) it is an integral part of human-centered and data-driven workflows~\cite{keimVisualAnalyticsDefinition2008}.
By utilizing computational methods and increasingly available machine-readable legal documents~\cite{vogelRichterKeinBot2024}, data-driven jurisprudence can efficiently extract knowledge from legal data, complementing traditional doctrinal analysis~\cite{lettierinicolaKnowledgeMachineriesIntroducing2019, vaciagogiuseppeOpportunitiesChallengesLegal2019}.
Natural Language Processing (NLP) and Artificial Intelligence (AI) have significantly advanced the efficiency and scalability of textual analysis in legal contexts, also driven by Large Language Models~(LLMs)~\cite{brownLanguageModelsAre2020}.
Nevertheless, the transition from linguistic processing to a computational understanding of legal language remains incomplete~\cite{baker2018LegalResearch2017}.
Current NLP techniques and AI models frequently lack the nuanced legal comprehension~\cite{vaciagogiuseppeOpportunitiesChallengesLegal2019} necessary to interpret legal material and infer legal concepts.

At the same time, the digitization of law is fraught with challenges. Substantial portions of legal documents remain restricted by license models and are rarely publicly available. Moreover, some countries lag in legislative efforts to make legal documents available in structured, digital formats~\cite{zanderUrteileUnterVerschluss2024}. In Germany, for example, less than 1\% of associated legal texts are openly available~\cite{ltoTransparenzJustizStagnation}.
Consequently, existing open German legal databases (refer to~\autoref{table:legal-document-collections}) offer limited functionality to improve efficiency. Interfaces are primarily \emph{keyword-based searches}, which often lack context-awareness and fail to account for implicit meanings or relationships~\cite{verlagc.h.beckohgBeckonline, jurisgmbhJuris, vogelRichterKeinBot2024}.

Despite these advancements, scientific efforts have primarily focused on computational linguistics and text-based analysis~\cite{Landthaler2016ExtendingFT, salaunConditionalAbstractiveSummarization2022, viannaOrganizingPortugueseLegal2022, correiaDynamicTopicModeling2023, bachingerExtractingLegalNorm2024}, leaving a gap in VA tailored to legal workflows.
Visualization and VA are critical for supporting knowledge work in fields such as digital humanities and healthcare~\cite{federicoRoleExplicitKnowledge2017}.
We understand the challenges of \textit{Visual Legal Analytics}~(VLA)~\cite{lettieriCartographiesLegalWorld2018} as a mismatch between traditional VA models and the workflows of legal scholars.
In legal contexts, traditional VA models are problematic since they often view knowledge as a singular entity resulting from user interactions with visualizations~\cite{keimVisualAnalyticsDefinition2008, sachaKnowledgeGenerationModel2014}. However, in jurisprudence, it can be insufficient to consider legal materials in isolation, requiring the application of domain knowledge. Hence, VA must distinguish between explicit, machine-interpretable knowledge, found in legal documents, and tacit knowledge, which is domain-specific, personal, and arises only from cognitive processing~\cite{wangDefiningApplyingKnowledge2009, federicoRoleExplicitKnowledge2017}.
In the legal context, tacit knowledge considers the knowledge about the inherent relationships between the laws and their application, which is often complicated to articulate~\cite{ruethersRechtstheorieUndJuristische2022}.

To address these unique needs of legal scholars, this work asks the following research question: \emph{In what ways can Visual Analytics support legal research by enhancing legal analytical reasoning?} It identifies previously unexamined opportunities but also challenges in digitalizing jurisprudence, thereby making the following contributions:
\begin{itemize}
    \item Empirical grounding through expert insights by conducting semi-structured interviews with nine legal practitioners to systematically identify domain-specific challenges and latent requirements in applying VA to jurisprudence.
    \item A three-phase workflow design that frames legal practice and generalizes to other fields of knowledge work, thereby fostering interdisciplinary methodology transfer.
    \item A case study proposing a VLA design to augment traditional interfaces, highlighting new opportunities in jurisprudence.
\end{itemize}

With these contributions, we provide a blueprint for a Visual Analytics design, considering tacit domain knowledge as essential for law applications and beyond.


\section{Background}
\label{section:background}

Jurisprudence is underpinned by methodologies and doctrinal analysis~\cite{bundesministeriumfuerjustizHandbuchRechtsfoermlichkeit2024}.
Although jurisprudence traditionally relies on such analysis, legal scholars increasingly recognize the importance of interdisciplinary research for addressing complex legal questions~\cite{siemsTAXONOMYINTERDISCIPLINARYLEGAL2009}.
For example, legal practitioners need to weigh up conflicting objectives as part of the European GDPR~\cite{gdpr2016} when considering legal aims and technical practicability regarding privacy and trust.

\subsection{Digital-Assisted Law Research}
Law research is primarily supported through commercial databases, like \beckOnline~\cite{verlagc.h.beckohgBeckonline} and \juris~\cite{jurisgmbhJuris} in Germany, similar to the Anglo-American \textit{Westlaw}~\cite{thomsonreutersWestlawClassic} and \textit{LexisNexis}~\cite{lexisnexislegal&professionalLexisNexis}. Those primarily feature keyword-search and backlinks.
\textit{LEX AI} offers tools for monitoring regulatory changes and curating personalized feeds with summaries and synopses of regulatory changes supported by AI~\cite{lexaigmbhLEXAI}.
\textit{\citeauthor{leretokgLeReToLegalResearch}} enriches documents with links to related materials~\cite{leretokgLeReToLegalResearch}. 
Meanwhile, tools like \textit{JURA KI Assistent}~\cite{ra-microsoftwareagJURAKIAssistent} and \textit{Justin Legal}~\cite{justinlegaltechgmbhJustinLegalDigitale2023} assist with tasks like legal drafting and management.
The European Union~(EU) offers several information retrieval tools for European and national law and jurisdiction, including EUR-Lex~\cite{publicationsofficeoftheeuropeanunionEURLex}, N-Lex~\cite{publicationsofficeoftheeuropeanunionNLex}, and the e-Justice Portal~\cite{directorate-generalforjusticeandconsumersEuropeanEJusticePortal}.
Furthermore, the EU provides LEOS~\cite{europeancommissionLEOS2015}, a collaborative online software platform for drafting European legislation with version control.
The EU advocates for the Interoperable Europe Act~\cite{theeuropeanparliamentandthecounciloftheeuropeanunionInteroperableEuropeAct2024}, which aims to enhance the interoperability of public administrations across its member states by implementing standardized technical solutions provided to EU citizens.
This act contributes to developing open legal standards, such as the European Legislation Identifier~\cite{publicationsofficeoftheeuropeanunionELI2017}.

\subsection{Visual Analytics for Law}

Close and distant reading~\cite{janickeVisualTextAnalysis2017} are well-established concepts in the Digital Humanities~(DH).
The distant reading paradigm quantifies and visually abstracts textual data as an integral part of discovery processes, offering new analytical insights~\cite{morettiGraphsMapsTrees2007, janickeVisualTextAnalysis2017}.
At the same time, Visual Analytics enhances understanding, promotes analytical reasoning, and aids decision-making in large information spaces using automated analysis techniques and interactive visual interfaces~\cite{keimVisualAnalyticsDefinition2008}.
This process is not systematic and thus requires tracking the provenance of the analysis steps and insights to review the hypotheses~\cite{shrinivasanSupportingAnalyticalReasoning2008, gotzCharacterizingUsersVisual2008}.
In favor of humans' limited working memory, a mechanism to externalize the evolving knowledge throughout the analysis process is required to establish links with the provenance~\cite{shrinivasanSupportingAnalyticalReasoning2008}.
Hence, VA is particularly promising for jurisprudence as it integrates human cognitive strengths into computational workflows.
For example, \citeauthor{tianLitVisVisualAnalytics2023} propose \textit{LitVis} for literature exploration and management that supports survey writing~\cite{tianLitVisVisualAnalytics2023}, conducting search, ranking, prioritization and classification.
\textit{KAMAS}~\cite{wagnerKnowledgeassistedVisualMalware2017} and \textit{VIStory}~\cite{zengVIStoryInteractiveStoryboard2021} adapt similar workflows to interactive storytelling, analytically reasoning about trends in scientific communities and the impact of highly-cited publications.
Together, these systems share key abstract tasks of information retrieval, understanding structural relationships, and analytical reasoning with jurisprudence~(for details, see \autoref{section:tasks}).
Nevertheless, systematic integration of VA into legal research workflows to support these tasks remains underexplored.
This encompasses (1) the dissemination of legal principles to the general public, (2) the comprehension of legal documents and the formulation of arguments, and (3) the assistance of legal professionals in comprehending legislation.
Thus, examining how VA can complement existing interdisciplinary methodologies in jurisprudence, particularly through interactive visualization and tacit knowledge, becomes essential.

\subsection{Legal Tech}

Legal technology, or legal tech, refers to all technological means that assist users in interacting with the law.
This includes the skills and techniques used to operate these means~\cite{whalenDefiningLegalTechnology2022}.
Importantly, this definition also encompasses information technology, such as visualization and AI.
Technological means can be categorized based on their directness and specificity. For instance, while a smartphone can be classified as legal tech, it is not limited to legal purposes and can require considerable effort to be used effectively in that context.
In contrast, practical research in legal tech focuses on designing applications specifically for legal needs, which require less input to function effectively.
For instance, \citeauthor{Landthaler2016ExtendingFT} improve semantic search in legal document collections using machine learning~(ML)~\cite{Landthaler2016ExtendingFT}.
Similarly, \citeauthor{viannaOrganizingPortugueseLegal2022} employ topic modeling, an ML technique, to organize Portuguese legal documents~\cite{viannaOrganizingPortugueseLegal2022}.
\citeauthor{bachingerExtractingLegalNorm2024} abstract specific applications into a workflow that automatically extracts entities from laws using LLMs~\cite{bachingerExtractingLegalNorm2024}.
To advance these computational methods, \citeauthor{darjiDatasetGermanLegal2023} present a dataset of legal norms annotated by legal experts~\cite{darjiDatasetGermanLegal2023}.
\citeauthor{fobbeCorpusDeutschenBundesrechts2024} also provides multiple digitized collections of German legal documents like the German federal law~\cite{fobbeCorpusDeutschenBundesrechts2024}.
In the legal community, visual aids such as a tabular representation~\cite{passalacquaUsingVisualTechniques1997}, mind maps, and flowcharts~\cite{khalilQuoVadisVisuelle2014, cantatoreMakingConnectionsIncorporating2016, mclachlanVisualisationLawLegal2021}, have been explored to communicate legal cases and convey legal concepts.
Additionally, other static visualizations enhance the comprehension of law~\cite{merklExplorationLegalText1997, katzMeasuringComplexityLaw2014, gomez2015understanding, ourednikVisualApproachHistory2017, bokwonleeNetworkStructureReveals2018}.
These include node-link diagrams, glyphs, geographic maps, and hierarchical representations.
\textit{EUCaseNet} and \textit{Knowlex} complement interactivity that enables information retrieval through visual exploration~\cite{lettieriComputationalApproachExperimental2016, lettieriLegalMacroscopeExperimenting2017}.
In addition, these approaches employ line graphs, heat maps, and tables.
\citeauthor{lacavaLawNetVizWebbasedSystem2022} support the understanding of structural relationships between legal documents in \textit{LawNet-Viz}~\cite{lacavaLawNetVizWebbasedSystem2022}.
\textit{LegalVis} enables analytical reasoning about judicial decisions through AI using bar charts and a text view of legal documents~\cite{resckLegalVisExploringInferring2023}.
These visualizations offer various perspectives on legal data.
While they aid legal scholars in navigating information spaces, they cannot review the analysis process or facilitate the externalization of knowledge crucial for effective legal reasoning.
However, such support is essential to uphold the essence of legal practice.
Moreover, merely presenting knowledge or relationships in a different format does not automatically guarantee benefits.
The provided capabilities must offer tangible advantages, be accepted, and then be utilized to maximize their transformative potential.
To better understand the opportunities that VA can offer for jurisprudence, we must, therefore, discuss legal research practice.

\section{Understanding Legal Research Practice}

Jurisprudence is a cornerstone of the legal domain, yet its complexity and reliance on vast information repositories pose significant challenges. Understanding legal research practices is essential to building effective applications with VA. Hence, we need to establish a common understanding of how legal scholars operate and which tasks are relevant to them. While jurisprudence is described extensively \cite{borchardJurisprudenceGermany1912, brugger1994legal, foster2010germanlaw, bundesministeriumfuerjustizHandbuchRechtsfoermlichkeit2024}, our work contributes new insights from the perspective of VA, incorporating existing knowledge and findings from the interviews with legal scholars.

\subsection{How Legal Scholars Operate - A Technical Perspective}
\label{section:how-legal-scholars-operate}

\begin{sidewaystable}
    \setcellgapes{.025 cm}
    \renewcommand{\arraystretch}{0.8}
    \makegapedcells
    \small
    
    \centering
    
    \caption{\textbf{A comparison of selected German legal document collections that are digitally available.}}
    \label{table:legal-document-collections}
    
    \begin{tabular}{r|ccccll}
        \toprule
        \multirow{2}{*}{\makecell[r]{\textbf{Name [Reference]}}} & \multicolumn{3}{c}{Type} & \multirow{2}{*}{\textbf{Open Access}} & \multirow{2}{*}{\makecell[l]{\textbf{Size} \\ (in documents)}} & \multirow{2}{*}{\textbf{Data Formats}} \\\cmidrule(lr){2-4}
	    & \textbf{\footnotesize Law} & \textbf{\footnotesize Commentary} & \textbf{\footnotesize Rulings} \\\hline
        openJur\textsuperscript{\textdagger}~\cite{openjurggmbhOpenJur} & $\blacksquare$ & $\square$ & $\blacksquare$ & $\LEFTcircle$ & $> 600,000$ & \texttt{\fontsize{8}{10}\selectfont HTML, PDF} \\
        Open Legal Data~\cite{ostendorffOpenPlatformLegal2020} & $\blacksquare$ & $\square$ & $\blacksquare$ & $\CIRCLE$ & $308,229$ & \texttt{\fontsize{8}{10}\selectfont HTML, JSON, XML} \\
        CDRS-BT~\cite{fobbeCorpusDrucksachenDeutschen2021} & $\blacksquare$ & $\square$ & $\square$ & $\CIRCLE$ & $131,835$ & \texttt{\fontsize{8}{10}\selectfont CSV, TXT, XML} \\
        CE-BGH~\cite{fobbeCorpusEntscheidungenBundesgerichtshofs2023} & $\square$ & $\square$ & $\blacksquare$ & $\CIRCLE$ & $77,892$ & \texttt{\fontsize{8}{10}\selectfont CSV, GraphML, PDF, TXT} \\
        CE-BPatG~\cite{fobbeCorpusEntscheidungenBundespatentgerichts2024} & $\square$ & $\square$ & $\blacksquare$ & $\CIRCLE$ & $30,866$ & \texttt{\fontsize{8}{10}\selectfont CSV, PDF, TXT} \\
        CE-BVerwG~\cite{fobbeCorpusEntscheidungenBundesverwaltungsgerichts2024} & $\square$ & $\square$ & $\blacksquare$ & $\CIRCLE$ & $27,200$ & \texttt{\fontsize{8}{10}\selectfont CSV, PDF, TXT} \\
        CE-BFH~\cite{fobbeCorpusEntscheidungenBundesfinanzhofs2023} & $\square$ & $\square$ & $\blacksquare$ & $\CIRCLE$ & $10,310$ & \texttt{\fontsize{8}{10}\selectfont CSV, HTML, PDF, TXT} \\
        CE-BVerfG~\cite{fobbeCorpusEntscheidungenBundesverfassungsgerichts2024} & $\square$ & $\square$ & $\blacksquare$ & $\CIRCLE$ & $8,949$ & \texttt{\fontsize{8}{10}\selectfont CSV, GraphML, HTML, PDF, TXT} \\
        Gesetze im Internet~\cite{bundesamtfuerjustizGesetzeImInternet} & $\blacksquare$ & $\square$ & $\square$ & $\CIRCLE$ & $6,800$ & \texttt{\fontsize{8}{10}\selectfont EPUB, HTML, PDF, XML} \\
        C-DBR~\cite{fobbeCorpusDeutschenBundesrechts2024} & $\blacksquare$ & $\square$ & $\square$ & $\CIRCLE$ & $6,784$ & \texttt{\fontsize{8}{10}\selectfont CSV, EPUB, PDF, TXT, XML} \\
        CE-BAG~\cite{fobbeCorpusEntscheidungenBundesarbeitsgerichts2020} & $\square$ & $\square$ & $\blacksquare$ & $\CIRCLE$ & $5,625$ & \texttt{\fontsize{8}{10}\selectfont CSV, PDF, TXT} \\
        Rechtsprechung im Internet~\cite{bundesamtfurjustizRechtsprechungImInternet} & $\square$ & $\square$ & $\blacksquare$ & $\CIRCLE$ & \textit{N/A} & \texttt{\fontsize{8}{10}\selectfont HTML, PDF, XML} \\
        OpinioIuris\textsuperscript{\textdagger}~\cite{shajkovciOpinioIuris} & $\blacksquare$ & $\blacksquare$ & $\blacksquare$ & $\LEFTcircle$ & \textit{N/A} & \texttt{\fontsize{8}{10}\selectfont HTML, PDF} \\
        Landesrecht BW~\cite{ministeriumdesinnerenfurdigitalisierungundkommunenBadenWurttembergLandesrecht} & $\blacksquare$ & $\square$ & $\blacksquare$ & $\LEFTcircle$ & $96,050$ & \texttt{\fontsize{8}{10}\selectfont HTML, PDF} \\\bottomrule
        beck-online\textsuperscript{\textdagger}~\cite{verlagc.h.beckohgBeckonline} & $\blacksquare$ & $\blacksquare$ & $\blacksquare$ & $\Circle$ & $> 55,000,000$ & \texttt{\fontsize{8}{10}\selectfont HTML, PDF} \\
        juris\textsuperscript{\textdagger}~\cite{jurisgmbhJuris} & $\blacksquare$ & $\blacksquare$ & $\blacksquare$ & $\Circle$ & $> 680,000$ & \texttt{\fontsize{8}{10}\selectfont HTML, PDF} \\
	\bottomrule
    \multicolumn{7}{@{}p{17.4cm}@{}}{{
    {\fontsize{6.2}{7}\selectfont The document types are marked with a filled square ($\blacksquare$), while an empty square ($\square$) signifies their absence. A filled circle ($\CIRCLE$) indicates document collections available in a structured, machine-readable format and eligible for automatic processing. In contrast, a half-filled circle ($\LEFTcircle$) designates the absence of one of these properties. Collections attributed with an empty circle ($\Circle$) are unavailable in a structured, machine-readable format and not eligible for automatic processing. Datasets superscripted with a dagger (\textsuperscript{\textdagger}) apply restrictive licensing prohibiting access to or processing their documents.\par}}
    }%
    \vspace*{-6mm}
    \end{tabular}
\end{sidewaystable}

The German civil law system builds upon a codified framework, anchored in the \textit{Grundgesetz für die Bundesrepublik Deutschland}~(Basic Law) and several codes, such as the \textit{Bürgerliches Gesetzbuch}~(Civil Code).
These codes form the foundation of the legal system and are supplemented by an extensive array of additional legal sources. 
Until the end of 2024, there are over 90,000 legal norms and regulations at the federal level, supplemented by numerous state and local statutes~\cite{mwo1797GesetzeUnd}.
Beyond codified laws, the system includes other essential sources such as \textit{explanatory memoranda}, \textit{legal commentaries}, \textit{court rulings}, and scholarly publications in \textit{trade journals} and \textit{commemorative volumes}~\cite[p.~52]{bundesministeriumfuerjustizHandbuchRechtsfoermlichkeit2024}.
Explanatory memoranda, published by the German parliament, elaborate on the legislative intent and provide interpretive guidance~\cite[p.~55]{bundesministeriumfuerjustizHandbuchRechtsfoermlichkeit2024}.
Legal commentaries, authored by scholars, interpret and analyze legal norms, while court rulings refine their application in practice~\cite[p.~52]{bundesministeriumfuerjustizHandbuchRechtsfoermlichkeit2024}. 
However, access to these sources is often restricted. In Germany, for example, legal commentaries are primarily available on the proprietary platforms \beckOnline~\cite{verlagc.h.beckohgBeckonline} and \juris~\cite{jurisgmbhJuris}, which impose licensing constraints.
Similarly, court rulings~--~particularly from lower courts~--~are frequently inaccessible due to data protection regulations.
While some publicly available datasets exist~(refer to \autoref{table:legal-document-collections}), they represent less than 1\% of legal decisions~\cite{ltoTransparenzJustizStagnation, vogelRichterKeinBot2024, zanderUrteileUnterVerschluss2024}.

\textbf{Systematic Relationships} ---
The interplay of these diverse legal sources constitutes the fabric of the German legal system~\cite[p.~460]{ruethersRechtstheorieUndJuristische2022}.
Understanding individual components requires knowledge of their relationships within the broader system.
These relationships are formally structured, hierarchical, and semantic, necessitating tacit knowledge to navigate their complexity effectively.
Applying legal sources follows a systematic process that relies on explicit procedures~\cite[p.~412]{ruethersRechtstheorieUndJuristische2022}.
Before applying legal norms, practitioners must first understand the sources that constitute the law.
Then, the application process typically unfolds through a series of distinct steps~\cite[p.~414]{ruethersRechtstheorieUndJuristische2022}.
First, practitioners establish the facts of the case, carefully analyzing the situation to identify relevant details.
Experienced practitioners often intuitively classify the case within a specific legal area or discipline~\cite[p.~413]{ruethersRechtstheorieUndJuristische2022}.
A challenging aspect of this process involves identifying the legal norms relevant to the case.
This task requires understanding the formally structured, hierarchical, and semantic relationships within the legal system~\cite[p.~460]{ruethersRechtstheorieUndJuristische2022}. 
Interpretation depends on contextualizing a norm within its broader legal framework, as its meaning and implications are frequently defined by its context.
Determining these connections relies heavily on the practitioner's expertise and familiarity.
Practitioners must continuously evaluate these relationships, considering both explicit legal provisions and implicit concepts derived from prior decisions and academic literature~\cite[p.~413]{ruethersRechtstheorieUndJuristische2022}.
After identifying the relevant norms, practitioners assess whether the case fulfills the specific elements outlined by these norms.
This process, known as \textit{subsumption}, involves comparing the facts with the requirements of the legal provisions~\cite[p.~413]{ruethersRechtstheorieUndJuristische2022}.
If the elements are satisfied, the practitioner determines the legal consequences accordingly.
However, cases are often not straightforward, and legal scholars can have differing opinions.

\subsection{High-Level Analysis Task Categories}
\label{section:tasks}

Legal tasks that span multiple levels of complexity, which can be categorized into the following, which we carefully base on the literature~\cite{bundesministeriumfuerjustizHandbuchRechtsfoermlichkeit2024, ruethersRechtstheorieUndJuristische2022, lacavaLawNetVizWebbasedSystem2022, lettieriLegalMacroscopeExperimenting2017, resckLegalVisExploringInferring2023, gomez2015understanding, mclachlanVisualisationLawLegal2021, burkhardtVisualLegalAnalytics2019} and our previous findings in \autoref{section:how-legal-scholars-operate}. For the process of distilling task categories, we orient ourself at the principles established in the HCI community~\cite{sedlmairDesignStudyMethodology2012, lamEmpiricalStudiesInformation2012, brehmerMultiLevelTypologyAbstract2013} to identifying, categorizing, grouping, and abstracting tasks:

\textbf{Fundamental Information Retrieval}~(\tasksFundamental) ---
Scholars focus on identifying and retrieving relevant legal sources at the foundational level, which includes filtering sources by topic~\cite{resckLegalVisExploringInferring2023}, understanding the formal legal organization, and determining specific legal norms within the broader legal corpus~\cite{lacavaLawNetVizWebbasedSystem2022}.
This includes downstream tasks such as \emph{finding the location of a legal norm}~\cite{lacavaLawNetVizWebbasedSystem2022}.

\textbf{Understanding Structural Relationships}~(\tasksRelationships) ---
Building on information retrieval, scholars analyze the structural relationships within legal texts, which involves evaluating the order of statutes~\cite{resckLegalVisExploringInferring2023} and interpreting both explicit and implicit relationships~\cite{lettieriLegalMacroscopeExperimenting2017, resckLegalVisExploringInferring2023}. Grasping these relationships is critical for understanding how individual legal norms interact within the broader legal framework.
This category includes downstream tasks such as \emph{recognizing a \emph{lex specialis} for a general law}.

\textbf{Advancing Legal Reasoning}~(\tasksReasoning) ---
At the highest level, legal reasoning tasks require synthesizing information to uncover patterns and identify conflicts across legal texts~\cite{gomez2015understanding, mclachlanVisualisationLawLegal2021}. Scholars engage in activities such as detecting inconsistencies between norms, identifying legal precedents, and formulating arguments. Additionally, they analyze legal conflicts~\cite{burkhardtVisualLegalAnalytics2019} and derive insights informing decision-making processes.
This includes downstream tasks such as \emph{reasoning about flaws and regulatory gaps}~\cite{ruethersRechtstheorieUndJuristische2022}.

\textbf{Specialized Tasks}~(\tasksSpecialized) ---
Beyond these general categories, certain tasks are unique to specific legal disciplines or application scenarios. Examples include \emph{Policy Modeling} such as analyzing the implications of proposed laws and policies~\cite{burkhardtVisualLegalAnalytics2019}, \emph{resolving natural language queries} through interpreting and answering complex legal queries~\cite{lacavaLawNetVizWebbasedSystem2022} or \emph{comparative law analysis}, which compares jurisdictions~\cite{lettieriLegalMacroscopeExperimenting2017}.

\section{Subject Matter Expert Interviews}
\label{section:interviews}
To deepen our understanding of the tasks performed by legal scholars and their associated challenges, we conducted semi-structured interviews with domain experts, discussing their established workflows and their opinions on potential technical support.

\subsection{Methodology}
\label{sec:methodology}

The semi-structured interviews with subject matter experts ($n = 9$) from the legal domain, averaging one hour each, followed a protocol but allowed for follow-ups to capture detailed insights. We conducted the interviews at the participants' workplace. Each interview began by gathering information about the participants' \emph{specializations and backgrounds}, asking them about their legal expertise, and professional and academic experiences.
Next, we inquired about the \emph{established workflows} for working with legal texts in specialization, including their tasks. The participants used their familiar setup to demonstrate how they carry out their workflows. We asked them to showcase these workflows using recent examples of their research. We also wanted the participants to describe the tools they use and their role in their workflows.
We also encouraged participants to \emph{compare their practices} in previous roles to establish diverse tasks.
Finally, we asked the experts to identify specific \emph{obstacles} they face with the tools they described, providing examples. Looking forward, we inquired about their \emph{needs and expectations} for tools that could improve their workflows.

\subsection{Participants}

The selected experts represents a diverse range of experiences: The participants included a professor of public law~($E_1$), seven post-doctoral~($E_{2}$) and doctoral~($E_{3-8}$) research associates specializing in various legal areas, and an undergraduate student of public management~($E_9$).
Concerning their experience, $E_1$ has experience in jurisprudence and legal practice, having spent 20 years in academia.
One set of participants focuses on German federal~($E_3$), state public~($E_4$), and European~($E_8$) law, all having practical experience in legal institutions~(e.g., public prosecutor offices, research services of the German parliament, European Commission, legal committees, the German ministry of the interior).
The remaining participants specialize in German civil law~($E_2$), administrative law~($E_5$), criminal law~($E_6$), private building law~($E_7$), while $E_9$ focuses on local politics and municipal law.

\subsection{Interviews}
\label{sec:interviews}

Legal scholars begin their workflow by considering all the facts about a specific case or research question.
Then, they identify laws and legal norms pertinent to their legal question.
As explained by $E_5$, the ``\textit{identification [...] is the first and simplest step}.''

Scholars follow a structured yet often non-linear approach for this task.
For example, $E_7$ succinctly summarizes their approach by identifying the law, understanding its structure, reading the reasoning, and finally delving into the jurisprudence and literature.
$E_3$, instead, initially conducts a keyword search to identify relevant legal norms.
Other participants~($E_2$, $E_4$, $E_5$, $E_7$, and $E_8$) concur and emphasize the significance of using keyword search tools.
All participants consistently refer to \beckOnline{} and \juris{} for this task, which they consider their primary literature database.
However, for European legislation, $E_3$ also mentions using Google Search, while $E_4$ refers to university document collections.
The variety of legal databases highlights a problem with the current user interfaces (UI), which are often incomplete silos that do not connect different databases, in particular when studying law across different legislative systems, hindering efficient workflows~($E_2$).
Some experts also wish for interface elements beyond the text that show interrelations and dependencies of laws that block specific processes~($E_1$), or how legal processes and administrative procedures follow step-by-step~($E_1$ and $E_7$).
Regarding \beckOnline{} and \juris{}, some scholars express concern, noting that ``\textit{keyword search is always quite poor, often with zero hits because [it is] too specific}''~($E_4$).
On the other hand, for too broad queries, $E_2$ frequently finds themselves manually screening through search results using their browser's search functionality.
$E_6$ also expresses dissatisfaction with the inefficiency of these keyword searches.
$E_3$, $E_4$, $E_8$, and $E_9$ highlight the manual aspect of their workflow when skimming through the table of contents of laws to locate potentially relevant legal norms.

At the same time, the participants acknowledge that ``\textit{many things are not in the text}'' ~($E_3$) but require ``\textit{the activation of implicit knowledge}''~($E_6$).
However, they regret that ``\textit{search engines do not have this context}''~($E_6$).
$E_4$ further emphasizes the interconnected nature of legal materials, highlighting their frequent reliance on explanatory memoranda, commentary, and court rulings that provide context for legal norms.
As $E_3$ confirms, ``\textit{detailed commentary is more interesting than shorter discussions}.''
$E_5$ also underscores the significance of legal materials linked to other laws and legal norms through references, which facilitate systematic interpretation.
As $E_6$ puts it, ``\textit{legal thinking is highly structured}.''
In light of this, $E_3$ and $E_7$ advocate for abstracting legal norms into concepts that enhance their research and facilitate compatibility with other legal systems.
For instance, $E_4$ employs concepts to compare court rulings that are not thematically related but contain legal ``\textit{reasoning and evidence}'' relevant to their discipline.
Furthermore, the meaning of terms can be defined elsewhere, leading to ``\textit{definition cascades}''~($E_3$) that require resolution through multiple layers and across different legal texts.
A related problem is the collision of disparate legal fields and competing legal definitions in intersecting areas ($E_1$, $E_9$), in particular, if criteria like \emph{lex specialis} (i.e., a specific law takes precedence over a general law) or \emph{lex posterior} (i.e., a newer law takes precedence over older legislation) cannot be applied, or a balancing of interests is required.
Despite these considerations, legal scholars must remain vigilant in wording, as $E_1$ emphasizes, ``\textit{three words change and the case looks completely different}.''
Further, the law only provides the guardrails and general pathways, as the legal interpretation allows for some discretionary decisions depending on the specific cases~($E_6$), for example, in agencies~($E_9$).
Legal scholars must bridge the gap between their domain knowledge and perspective, also communicating with their stakeholders, including employees, managers, politicians, and the general public, in their interpretation~($E_1$).

In the face of these requirements, participants remain skeptical about digitization and LLMs in the legal domain.
First and foremost, $E_8$ continues to identify ``\textit{digitization as a problem}'' within the legal domain.
For instance, electronic files in courts still encounter practical challenges~($E_8$), and court rulings are often not digitally accessible~($E_4$) or follow restrictive copyright and licensing models ($E_6)$.
$E_4$ further highlights that these inaccessible documents disrupt their workflow, as they must scan them before proceeding with their research.
Regarding LLMs, $E_5$ finds that using GPT still requires a lot of trial and error to achieve satisfactory results when connected with search engines.
$E_2$ agrees that LLMs ``\textit{lack traceability}'' but still believes that AI could be relevant in supporting whether a text is relevant and worthwhile to consider during legal research.
However, participants still prefer to read them in full themselves, which $E_8$ also finds to facilitate serendipitous discovery of relevant legal norms and arguments.
$E_7$ occasionally uses LLMs for mundane tasks only tangibly to legal research questions, such as preparing interviews.

\section{Addressing Challenges in Legal Research: A Three-Phase Workflow Design}
\label{section:workflow}
\begin{figure}
	\includegraphics[width=\linewidth]{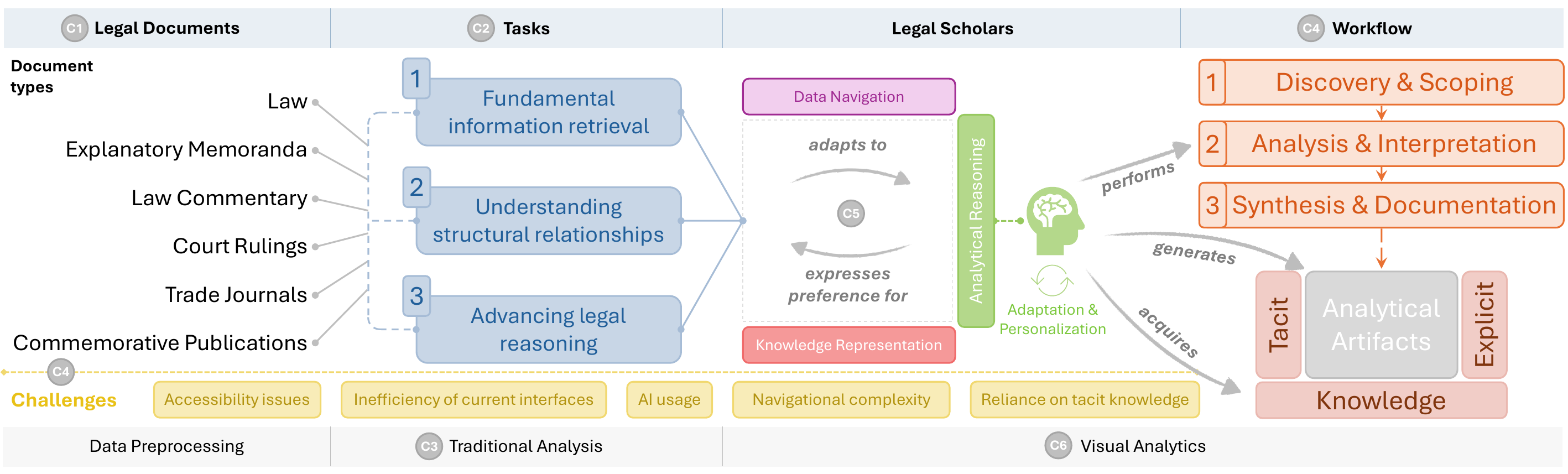}
	\centering
    \caption{\textbf{Using Visual Analytics (VA) for legal research} can simplify working with diverse legal documents~--~including laws, explanatory memoranda, and court rulings~--~to support legal scholars in three key tasks: (1) fundamental information retrieval, (2) understanding structural relationships, and (3) advancing legal reasoning, while addressing several challenges.
    Through interviews with domain experts, we identify three workflow phases, namely (1) Discovery \& Scoping, (2) Analysis \& Interpretation, and (3) Synthesis \& Documentation. Through them, scholars iteratively navigate through and generate analytical artifacts that capture explicit and tacit knowledge, fostering a deeper comprehension of legal structures and reasoning, and informing a VA design for jurisprudence.}
    \label{figure:teaser}
\end{figure}

In line with these insights, and building on the bottom-up findings from our subject-matter expert interviews, we observe that legal scholars follow structured steps in their research~\cite[pp.~412--413]{ruethersRechtstheorieUndJuristische2022}.
Most participants begin by conducting a keyword search to identify relevant laws and legal norms.
However, a single attempt is usually insufficient for satisfactory results, so keyword search becomes an iterative process~\cite[p.~413]{ruethersRechtstheorieUndJuristische2022}.
Often, key insights emerge not through targeted searches but during comprehensive reading.
Subsequently, legal scholars transition to contextualizing laws and legal norms with each other, using commentary and other legal materials~\cite[pp.~93, 459--461]{ruethersRechtstheorieUndJuristische2022}.
This involves connecting the initially narrow scope of documents with a broader landscape of legal materials to enable systematic interpretation.
Given the vast legal document collections, this process can be cognitively taxing for scholars.
Therefore, they rely on documentation throughout the process~\cite{cantatoreMakingConnectionsIncorporating2016}.
Documenting their legal insights enables synthesis, which is used to construct legal arguments and constitute legal knowledge.
Based on these observations and the literature, we identify a \emph{three-phase} workflow~(\workflowFirstPhase - \workflowThirdPhase)~(see also \autoref{figure:teaser}). The phases that we define~--~Discovery \& Scoping~(\tasksFundamental), Analysis \& Interpretation~(\tasksRelationships), and Synthesis \& Documentation~(\tasksReasoning)~--~illustrate the systematic but manual processes scholars typically follow.
To illustrate, consider \textit{Alice}, a legal scholar with extensive experience in German legislation.
She is researching racial profiling in the context of the legal safeguards offered by German law against law enforcement from engaging in racial profiling.
The following three sections describe the workflow's phases as executed \emph{traditionally} without VA, outlining legal scholars' challenges, where $E_i$ refers to an exemplary experience described by a participant in~\autoref{section:interviews}.

\subsection{\workflowFirstPhase: Discovery \& Scoping}
\label{section:workflow-first-phase}

In the first phase of the workflow (\workflowFirstPhase), Alice identifies suitable legal texts and narrows them down to address her research problem~($E_7$;~\tasksFundamental).
Familiar with the commercial legal database \beckOnline~\cite{verlagc.h.beckohgBeckonline}, she begins by querying the platform's search interface.
Alice knows from recurring media reports that cases of racial profiling are sometimes prevalent in operations conducted by the \textit{Bundespolizei} (German Federal Police), whose responsibilities include securing German borders~(\S{}~2~BPolG~(Federal Police Act)), train stations~(\S{}~3~BPolG), and airports~(\S{}~4~BPolG)~(\textit{\challengeReliance{}}).
She starts with the search phrase ``\textit{Racial Profiling Bundespolizei}.''
Her query yields only five matching documents, prompting Alice to broaden her search~($E_8$) using the more general phrase ``\textit{Racial Profiling}.''
While this query produces a significantly larger volume of results, it also introduces many irrelevant entries~(\textit{\challengeInefficiency{}}).
Alice begins her review by examining the most relevant legal norms by \beckOnline{} for her search queries.
These include Art.~3~GG~(Basic Law), \S{}~23~BPolG, and Art.~14~EMRK~(European Convention on Human Rights).

Next, Alice evaluates the documents in the search results~($E_2$).
The results are sorted by relevance, but the platform does not make the ranking criteria transparent, leaving Alice uncertain about the prioritization~(\textit{\challengeInefficiency{}}).
She skims through the initial results and discovers that some relevant documents are inaccessible due to the limits of her subscription~(\textit{\challengeAccessibility{}}).
The first results predominantly include journal articles, court rulings, and commentaries rather than primary legal statutes.
To supplement this, Alice manually expands her search to explanatory memoranda in other databases~($E_4$), such as the \emph{Dokumentations- und Informationssystem für Parlamentsmaterialien}~(DIP)~\cite{deutscherbundestagDIPDokumentationsUnd}~(\textit{\challengeComplexity{}}).
Recognizing the importance of these documents for contextual analysis, she notes them for detailed inspection in~\workflowSecondPhase.

\subsection{\workflowSecondPhase: Analysis \& Interpretation}
\label{section:workflow-second-phase}

In the second phase (\workflowSecondPhase), Alice analyzes the documents identified in the first phase~\workflowFirstPhase~(\tasksRelationships).
She focuses on connecting the legal norms highlighted by \beckOnline{}~--~such as Art.~3~GG, \S{}~23~BPolG, and Art.~14~EMRK~--~with the contents of the selected documents~($E_3$).
\beckOnline{} allows Alice to preview these norms, but does not support visualizing relationships between them.
Alice draws on her expertise to recall that \S{}~23~BPolG, which permits identity checks by the federal police, is conceptually linked to \S\S{}~2-4~BPolG, which outline the responsibilities of the federal police~(\textit{\challengeReliance{}}).
However, \beckOnline{} lacks tools to map or visualize such connections explicitly, requiring Alice to track the relationships as she progresses through her analysis mentally~(\textit{\challengeComplexity{}}).

She begins by examining the first matching document from a journal.
Due to its limited structural organization, Alice skims the text to identify relevant legal norms and notes them mentally~($E_2$). 
This process repeats for the remaining documents, including those accessed through supplemental searches in other databases.
Alice consults their details whenever encountering unfamiliar norms by following backlinks in the legal database.
Through her analysis, Alice identifies arguments supporting her assumption that German law inadequately prevents racial profiling. 
However, eliciting these insights requires substantial manual effort as she digs through multiple documents.
Alice also suspects that norms related to the state police laws (such as the \textit{Polizeiaufgabengesetz}~(Police Tasks Act) in Bavaria) might offer additional insights.
However, she finds no references to these in her current search results, indicating a gap in the coverage~(\textit{\challengeAccessibility{}}).

\subsection{\workflowThirdPhase: Synthesis \& Documentation}
\label{section:workflow-third-phase}

In the final phase (\workflowThirdPhase), Alice synthesizes her findings and documents the outcomes of her analysis~($E_4$;~\tasksReasoning).
She integrates her mental map of relationships between legal norms with the arguments and insights derived from the reviewed documents.
She organizes these findings into a coherent narrative using a word processing program, ensuring that references to relevant legal texts and scholarly arguments support her conclusions.
Alice imports the key documents into her reference management software for citation and further reference~($E_2$).
This phase remains labor-intensive due to the absence of automated insights from multiple sources that visualize the relationships between legal norms~(\textit{\challengeComplexity{}}).
Despite these limitations, Alice completes the synthesis and prepares to revisit it in subsequent iterations, aiming to address gaps identified in previous phases.

\section{Visual Analytics for Jurisprudence}
\label{section:visual-analytics-for-jurisprudence}

We group the interviewees’ comments into recurring \emph{challenges} faced by legal scholars when analyzing and applying legal materials:

\begin{enumerate}
    \item \textbf{\challengeAccessibility{}:} Limited digital access and licensing restrictions hinder the use of commentaries and rulings, especially from lower courts. Participants ($E_8$) stressed that \emph{data availability} accelerates research~(\tasksFundamental), while relying on print sources causes delays~($E_5$). They also noted that the duopoly of \beckOnline{} and \juris{} restricts access, and that lower-court rulings are often inaccessible due to data protection rules~\cite{ltoTransparenzJustizStagnation}.

    \item \textbf{\challengeComplexity{}:} Navigating relationships between legal texts requires substantial manual effort. Participants reported using tables of contents~($E_7$, $E_9$), explanatory memoranda~($E_4$, $E_7$), or manual searches when links were missing~($E_2$). They emphasized the need for tools that better support exploratory \emph{navigation} of document hierarchies and semantic relationships~(\tasksRelationships).

    \item \textbf{\challengeInefficiency{}:} Existing platforms (e.g., \beckOnline{}, \juris{}) emphasize syntactic rather than semantic search. Participants cited limited support for domain-specific language~($E_4$, $E_6$, $E_7$), failure to find implicit legal concepts~($E_3$), and over-reliance on keyword matching~($E_4$;~\tasksReasoning). To cope, some ($E_2$, $E_4$, $E_5$) used browser search within results, underscoring the demand for features such as fuzzy or semantic search and contextualization~(\tasksSpecialized, $E_6$).

    \item \textbf{\challengeAI{}:} Interviewees showed cautious interest in \emph{AI} tools, stressing the need for reliability and transparency. While $E_6$ rarely used \emph{LLMs}, $E_2$ and $E_7$ raised concerns about their trustworthiness. $E_2$ reported using LLMs to filter search results~(\tasksFundamental), suggesting guarded optimism about their potential.

    \item \textbf{\challengeReliance{}:} Legal reasoning depends on relationships between documents, but many links remain implicit. Scholars rely on their domain expertise to infer connections, a process they described as labor-intensive and error-prone. They identified value in tools that could better integrate explicit and tacit knowledge, which is currently labor-intensive and error-prone.
\end{enumerate}

\noindent Apart from open data, these findings emphasize the need for more adequate tooling support in the legal domain.
The individual phases in the workflow~(see \autoref{section:workflow}) need different types of support for these challenges.
As discussed before~(see \autoref{section:background}), some of the challenges align with standard practices in text-based research domains, which allows us to reuse some existing concepts, but with the need to adapt and integrate them into the specific needs of law scholars.

\begin{figure*}
    \centering
    \includegraphics[width=\linewidth]{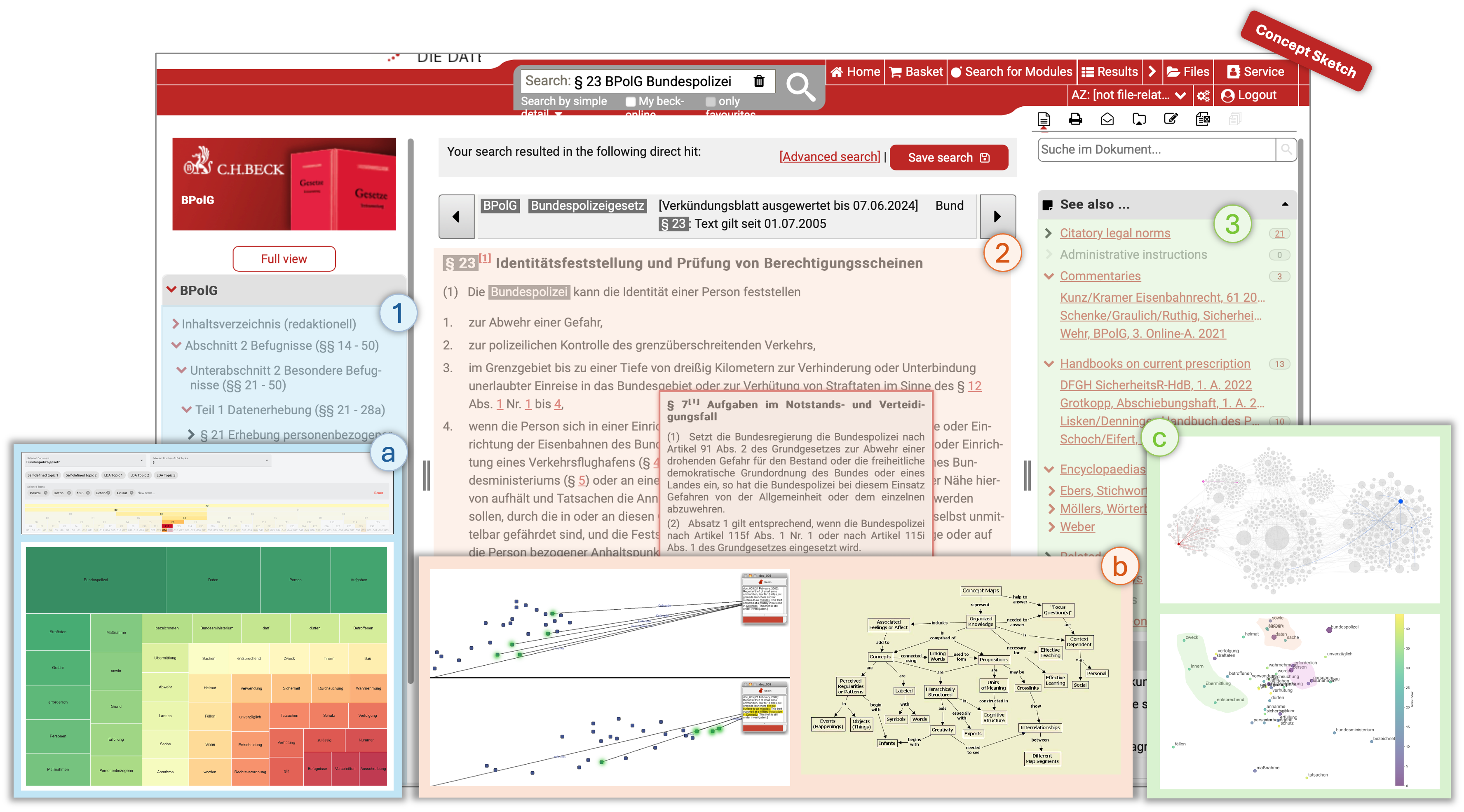}
    
    \caption{\textbf{An annotated concept sketch of a legal database's user interface that displays the results for a search query contextualized with VA techniques.} On the left-hand side, the user interface~(1)~displays the formal legal structure of the \emph{Bundespolizeigesetz} (Federal Police Act). At the center, the database~(2)~renders the selected legal norm's text with keywords from the query highlighted. On the right-hand side, the user interface~(3)~lists related documents such as legal norms and law commentaries.}
    \label{figure:user-interface-overview}
\end{figure*}

We identify three key areas of VA that hold significant benefits for jurisprudence: 
(1)~\emph{data navigation}: for effective exploration and cross-connection of document collections, (2)~\emph{knowledge representation}: for properly structuring data and metadata while integrating domain knowledge with explicit knowledge from legal entities, and (3)~\emph{analytical reasoning}: for synthesizing and documenting findings with the support of AI.
In \autoref{figure:user-interface-overview}, we showcase the visual components of an exemplary \emph{traditional user interface} of a commonly used German legal database.
Alongside this traditional UI, we propose potential enhancements through VA techniques.
The traditional keyword-based search (see \autoref{section:interviews}) consists of three parts: (1)~the formal legal structure of the inspected law for data navigation as a hierarchical list, (2)~the display of the legal norm's contents as a main body of text, and (3)~a simple list of related legal documents.
Based on the literature and our findings, the following enhancements are possible.
On the left-hand side, in addition to the hierarchical list, the UI (1) visualizes the hierarchy of the \emph{Bundespolizeigesetz}~(Federal Police Act) using a collapsible tree structure as a conceptual tree map.
It contains the law's formal structure labeled by its section titles.
Users can navigate the law through interaction with the structure, moving between legal norms.
The UI (2) renders the selected legal norm's contents at the center, representing explicit knowledge.
The rendering highlights the keywords from the query in the contents verbatim.
The user interface backlinks to their documents for explicitly referenced legal entities, simultaneously displaying a preview on hover.
On the right-hand side, the legal database (3) lists legal documents related to the currently displayed document, like legal norms, law commentary, and court rulings.
For the three key areas of VA discussed above, we analyze the deficiencies, explore the state-of-the-art literature, and distill possible improvements in jurisprudence through VA, while also discussing the challenges at this intersection.

\subsection{Data Navigation}
\label{section:va_data_navigation}

In VA, data navigation encompasses the exploration of and the movement in information spaces~\cite{shrinivasanSupportingAnalyticalReasoning2008}.
For jurisprudence, this translates to navigating formally structured legal document collections.
The traditional UI of legal databases (refer to \autoref{figure:user-interface-overview}) relies on a collapsible tree-based list to represent hierarchies.
However, this approach fails to provide semantic context during navigation.
Since legal structures can be nested several levels deep~\cite[p.~152]{bundesministeriumfuerjustizHandbuchRechtsfoermlichkeit2024}, their tree representation can become cumbersome.
Additionally, tree-based lists inadequately support the simultaneous navigation of multiple legal structures~--~a common requirement in jurisprudential research.

\citeauthor{merklExplorationLegalText1997} introduced the concept of navigating such corpora using a \textit{Hierarchical Feature Map}~\cite{merklExplorationLegalText1997}, a Treemap variant that organizes clusters across layers representing the hierarchy.
Each layer corresponds to a hierarchical level, with subsequent layers revealing sub-clusters.
Despite their improvement, the static, semi-automatically generated maps require manual input for cluster titles, limiting scalability.
Building on this idea, \citeauthor{lettieriLegalMacroscopeExperimenting2017} started employing Treemaps~\cite{johnsonTreeMapsSpaceFillingApproach1991} in the \textit{Legal Doctrine Semantic Navigator} for navigating formally structured legal documents through drill-down and roll-up operations~\cite{lettieriLegalMacroscopeExperimenting2017}. 
Later, they introduced the \textit{sliding Treemap}, a variation, enabling the navigation of rulings by the European Court of Justice and other legal documents on mobile devices~\cite{lettieriAffordanceLawSliding2020}.
Treemaps with interactive drill-downs can display multiple paths simultaneously across hierarchies and can visually encode guidance by enlarging documents of higher relevance~\cite{lettieriLegalMacroscopeExperimenting2017}~(refer to \autoref{figure:user-interface-overview}, a).

Beyond Treemaps, Icicle Plots~\cite{kruskalIciclePlotsBetter1983} have emerged as an effective visualization method for hierarchical data.
DH researchers employ these plots for document comparison~\cite{tytarenkoHierarchicalTopicMaps2024}.
\citeauthor{kimArchiTextInteractiveHierarchical2020} use Icicle Plots to visualize a hierarchy of topics during their interactive modeling~\cite{kimArchiTextInteractiveHierarchical2020}, while \citeauthor{tytarenkoHierarchicalTopicMaps2024} also visualize a hierarchy of topics to compare documents~\cite{tytarenkoHierarchicalTopicMaps2024}~(refer to \autoref{figure:user-interface-overview}, a).

We propose a hybrid approach combining the strengths of both visualization techniques based on the following observations.
Since Treemaps perform worse at displaying large hierarchies than Icicle Plots~\cite{macquistenEvaluationHierarchicalVisualization2020}, distributing the visual representation between these complementary visualizations is beneficial.
A Treemap would display the first few layers of the currently selected path as an overview, while a juxtaposed Icicle Plot would present the remainder of the path in the hierarchy.
Employing the two visualizations in parallel mitigates scalability issues and leverages the benefits of interactively linking both visualizations~\cite{burchPowerInteractivelyLinked2022}, offering complementary perspectives on legal corpora.
This combination would support legal scholars during the initial workflow phase~(refer to \hyperref[section:workflow-first-phase]{\workflowFirstPhase}), facilitating the discovery and scoping of legal document collections.

Alternative hierarchical visualization techniques, such as Bubble Treemaps or Sunbursts~\cite{macquistenEvaluationHierarchicalVisualization2020} can also be employed but perform worse for small hierarchies than our proposed approaches.
Still, we emphasize the need for exploring other design alternatives such as circular grid layouts~\cite{qiuVADISVisualAnalytics2025} and hierarchical edge bundling~\cite{holtenHierarchicalEdgeBundles2006} for radial visualizations like a radial dendrogram.
Though less prevalent in the legal domain, these alternatives are worth considering for expanding the design space.

\subsection{Knowledge Representation}
\label{section:va_knowledge_representation}

The interactions with visualizations for data navigation heavily depend on the user's prior knowledge, explicit and tacit~\cite{wangDefiningApplyingKnowledge2009}.
In jurisprudence, explicit knowledge can manifest as connections between legal documents~\cite{lacavaLawNetVizWebbasedSystem2022}, where one norm references another legal norm, establishing a relationship.
The traditional UI of legal databases displays these relationships through backlinks, with a pop-up and metadata, thereby representing explicit knowledge.
The tree-based list of the law's hierarchy~(left) and the list of related legal documents~(right) also express explicit knowledge contained in legal documents.
Meanwhile, legal scholars' tacit knowledge about the norm application is internalized as part of legal scholarship. However, it is not articulated~\cite{ruethersRechtstheorieUndJuristische2022}, which would be especially important for legal reasoning~(see \autoref{section:workflow-second-phase}).

The literature proposes visualizations such as stacked graphs and geographic maps, which have been employed to visualize the metadata of legal documents~\cite{gomez2015understanding}.
Similarly, \citeauthor{resckLegalVisExploringInferring2023} visualize legal structure by semantically segmenting it by paragraphs~\cite{resckLegalVisExploringInferring2023}.
Meanwhile, other works statically analyze the complexity by visualizing explicit references between articles~\cite{bokwonleeNetworkStructureReveals2018}.
Beyond jurisprudence, explicit knowledge of text-based documents is visualized in the context of discussion forum posts~\cite{jian-syuanwongMessageLensVisualAnalytics2018} and document comparison~\cite{tytarenkoHierarchicalTopicMaps2024}.
However, these examples often neglect to facilitate the articulation of tacit knowledge.
Instead, VA should consider tacit knowledge~\cite{federicoRoleExplicitKnowledge2017}.
For instance, \citeauthor{wagnerKnowledgeassistedVisualMalware2017} allow cybersecurity experts to classify malware behavioral rules~\cite{wagnerKnowledgeassistedVisualMalware2017}, while \citeauthor{mistelbauerSmartSuperViews2012} enhance medical visualizations by ranking views based on experience~\cite{mistelbauerSmartSuperViews2012}.
\citeauthor{endertSemanticInteractionVisual2012} define several semantic interactions that support analytical reasoning for text analytics~\cite{endertSemanticInteractionVisual2012}.
The authors also provide an implementation, letting users determine the importance of phrases through highlighting.
Accordingly, documents containing the exact phrase become closer in a spatial layout~(refer to \autoref{figure:user-interface-overview}, b).
Similarly, a diagram of concepts connected by labeled arcs constitutes a \emph{Concept Map}, where the arcs created by semantic interactions describe tacit knowledge~\cite{canasConceptMapsIntegrating2005}~(refer to \autoref{figure:user-interface-overview}, b).

Translating these solutions from VA to jurisprudence, we observe that visually laying out legal concepts in a map is an established approach~\cite{burkhardtVisualLegalAnalytics2019, ginevraperuginelliKnowledgeLawBig2019, onamiLegalVizLegalText2025}.
Concept maps can enhance VA for jurisprudence by converting legal scholars' domain knowledge into explicit knowledge through semantic interactions, thereby improving the effectiveness of LLMs.
With \textit{KMTLabeler}, \citeauthor{wangKMTLabelerInteractiveKnowledgeAssisted2024} propose a labeling tool for medical texts integrating VA with LLMs~\cite{wangKMTLabelerInteractiveKnowledgeAssisted2024}, allowing experts to express their domain knowledge by keyword labeling rules.
We can also connect the knowledge representation to legal reasoning since a \emph{Concept Map} resembles a network.
Both support the latter two phases of the jurisprudence workflow~(refer to \hyperref[section:workflow-second-phase]{\workflowSecondPhase} and \hyperref[section:workflow-third-phase]{\workflowThirdPhase}).

While networks are suitable for supporting this workflow phase, node-link diagrams may not be the most effective visualization method. Matrices, i.e., tabular displays of graphs, are proven methods for visualizing network data, but are underutilized in VLA. Hence, studying hybrid approaches and alternative representations specific to the legal domain can enrich the VLA design space.

\subsection{Analytical Reasoning}
\label{section:va_analytical_reasoning}

The traditional UI of legal databases merely lists groups of legal documents related to the currently displayed document.
There are no means by which legal scholars can add evidence of their insights to that list.
In VA, analytical reasoning is a knowledge-generation process that can have systematic aspects but is also serendipitous~\cite{shrinivasanSupportingAnalyticalReasoning2008}.
Users must be aware of the evidence found throughout this process to increase the chance of insights.
A prominent approach to capture such evidence is analytical provenance as suggested by \citeauthor{pirolli2005sensemaking} for their sense-making loop~\cite{pirolli2005sensemaking}.
For that, \citeauthor{perez-messinaPersistentInteractionUserGenerated2024} suggest organizing analytical artifacts like annotations and their relationships~\cite{shrinivasanSupportingAnalyticalReasoning2008, perez-messinaPersistentInteractionUserGenerated2024}.

In related text-based research practice, \textit{Overview} allows investigative journalists to create tags that attach their findings to clusters~\cite{brehmerOverviewDesignAdoption2014}.
\citeauthor{tianLitVisVisualAnalytics2023} add to this, allowing users to adapt clusters of literature interactively, adding comments about their understanding of topics~\cite{tianLitVisVisualAnalytics2023}~(refer to \autoref{figure:user-interface-overview}, c).
\textit{ForceSPIRE} addresses visual text analytics more broadly, enabling experts to create text highlights that constitute a spatial layout of document relationships~\cite{endertSemanticInteractionVisual2012}~(refer to \autoref{figure:user-interface-overview}, b).

Conveying the relationships within and between such clusters is central for analytical reasoning, so node-link diagrams are frequently employed as visualization methods.
\citeauthor{lettieriLegalMacroscopeExperimenting2017} introduce the \textit{Norm Graph Navigator}, allowing interactive exploration of legal relationships~\cite{lettieriLegalMacroscopeExperimenting2017}.
More recently, \citeauthor{lacavaLawNetVizWebbasedSystem2022} integrate natural language querying into network navigation, enabling users to filter nodes efficiently~\cite{lacavaLawNetVizWebbasedSystem2022}~(refer to \autoref{figure:user-interface-overview}, c).
Since analytical artifacts externalize tacit knowledge as explicit~\cite{wangDefiningApplyingKnowledge2009}, they can act as a countermeasure to the confirmation bias in hypothesis generation~\cite{pirolli2005sensemaking} and enable collaboration~\cite{perez-messinaPersistentInteractionUserGenerated2024}, which can reinforce diverse opinions in legal decision-making.
When persisted~\cite{perez-messinaPersistentInteractionUserGenerated2024}, these artifacts can help users recall and replicate insights~\cite{raganCharacterizingProvenanceVisualization2016}.
Analytical provenance additionally fosters transparency, which is particularly important in the legal domain, where the validity and reliability of insights must be meticulously documented.
Integrating these aspects, for example, through 2D clustering for overview, graph networks for relations, and provenance trees for documentation, shows the significant potential of VA for jurisprudence.

\section{Emergent Challenges for Visual Analytics Design}
While numerous approaches exist for data navigation in law, knowledge representation and reasoning are often left to analysts to perform manually, as we have seen in the previous discussion.
Crucially, no existing solution unifies all these areas of VA into a framework that exploits their synergetic effects.
In the following, we explore the \emph{opportunities and challenges} of developing such an integrated approach.

\subsection{Addressing Jurisprudence}
\label{section:addressing-jurisprudence}

\begin{figure}[!tbhp]
    \includegraphics[trim={0 1.5cm 0 1.8cm},clip, width=\linewidth]{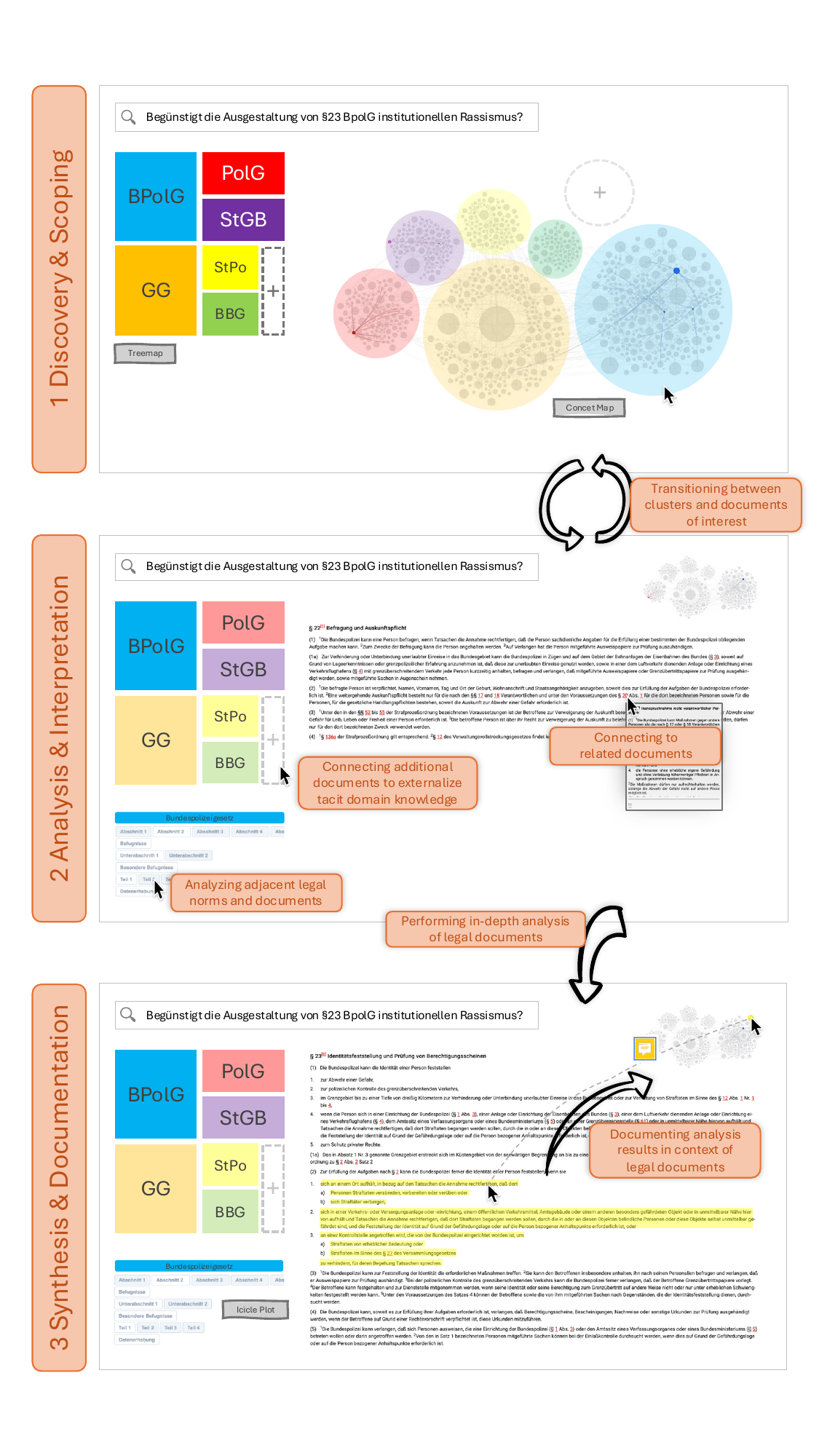}
    \centering
    \caption{A mock-up of our proposed Visual Analytics workflow that illustrates the role of Concept Maps, Icicle Plots, and Treemaps in the different phases of the three-phase workflow. Cursors and arrows indicate possible user interactions.}
    \label{figure:mock-up}
\end{figure}

Traditionally, German legal scholars predominantly use keyword-based search interfaces, the prevalent UI paradigm~\cite{verlagc.h.beckohgBeckonline, jurisgmbhJuris}.
However, this poses several challenges to legal scholars, hindering effective workflow support.
While familiar, the paradigm limits scholars to primitive input, lacking the expressiveness required in the legal domain.
Hence, we propose combining existing visualization techniques into a novel VA workflow~(refer to~\autoref{figure:teaser}) for jurisprudence with LLMs.

As part of the initial workflow phase~(refer to \hyperref[section:workflow-first-phase]{\workflowFirstPhase}), the scholar scopes their research verbally. Instead of a syntactic keyword search, they articulate analysis goals using natural language~(refer to~\autoref{figure:mock-up}). The system processes the verbalization and embeds the query into a semantic context by inferring references to legal entities such as laws, legal norms, and court rulings. We imagine an overview displaying clusters of legal concepts and individual documents relevant to the query, enabling scholars to leverage their domain expertise to refine the selection. For its implementation, the overview leverages a network visualization augmented with colored shapes that indicate clusters. For details, a scholar can preview individual documents on demand during selection, highlighting explanations for their relevance.
After refining the scoping, legal scholars transition to the second workflow phase~(refer to \hyperref[section:workflow-second-phase]{\workflowSecondPhase}). In VA, the UI transitions into a detailed view upon selecting a cluster or document of interest~(refer to~\autoref{figure:mock-up}). We propose juxtaposing the formal legal structure with a branching visualization such as an Icicle Plot. The former displays the legal documents of the scope with their hierarchy, enabling efficient navigation. Meanwhile, the Icicle Plot provides the contextualized path of the selected document in its hierarchy. Together, these visualizations address the legal domain's unique requirement for simultaneous navigation across multiple hierarchies.

During the analysis and interpretation, legal reasoning takes place, which must utilize the scholar's domain knowledge. While the proposed VA techniques can initially visualize the explicit knowledge from the legal documents, they cannot draw on evolving domain knowledge. To achieve this, VA must provide interaction for users to articulate their tacit knowledge. Therefore, at any point during the workflow, if users find a relevant document missing from the current selection, they can establish a relationship to include it. The Treemap and Icicle Plot visualizations facilitate this interaction by integrating empty cells with suggestive icons as visual placeholders. By clicking these icons, the analyst can relate a selected document to others represented in the visualizations. These interactions reveal user preferences relevant to the subsequent recommendation of related legal documents. With these user preferences, we augment a knowledge graph (KG) of the legal documents selected during the first workflow phase. \citeauthor{yangLegalGNNLegalInformation2021} demonstrate the benefit of such an augmentation with \textit{LegalGNN}, a framework for recommending legal documents based on a graph neural network trained with a legal KG and the contents of legal documents~\cite{yangLegalGNNLegalInformation2021}. We exploit the resulting recommendations to personalize the VA workflow with the expert's domain knowledge. Further, we draw tacit knowledge from its externalization through analytical artifacts.

Besides the Treemap and the Icicle Plot, we propose to add a concept map to the top side of the suggested user interface~(refer to~\autoref{figure:mock-up}). The map allows users to synthesize and document evidence acquired throughout the legal reasoning process, forming a crucial part of the third workflow phase (refer to \hyperref[section:workflow-third-phase]{\workflowThirdPhase}). Scholars capture and connect relevant artifacts within the concept map to organize their legal insights. Nodes represent primitive artifacts resembling individual legal entities, such as legal norms or court rulings. In contrast, compound artifacts describe the links that relate multiple primitives, representing causal connections in the legal realm. Based on these and the graph augmented during the previous workflow phase, we enhance documentation of the legal reasoning process through retrieval-augmentation generation (RAG). \citeauthor{edgeLocalGlobalGraph2025} propose \textit{GraphRAG}, which performs RAG on the KG of text corpora~\cite{edgeLocalGlobalGraph2025}. The authors leverage the hierarchy of the KG to enable the summarization of documents at varying levels of abstraction. This approach aligns with the nature of legal corpora and can benefit the documentation of legal reasoning in jurisprudence. For instance, in analyzing racial profiling by the German Federal Police~(see \autoref{section:workflow}), a scholar might document that the police are responsible for securing train stations~(\S{}~3~BPolG) and are permitted to identify individuals if necessary for fulfilling their duties~(\S{}~23~BPolG). However, this authority must be balanced against the third-party effect of the German Basic Law~(Art.~3~GG), which ensures equality and non-discrimination.

While the concept map functions as an isolated visualization, we propose a novel augmentation that merges it with the network visualization used during the first workflow phase~(refer to \hyperref[section:workflow-first-phase]{\workflowFirstPhase}). This integration creates a seamless transition between overview and detail, addressing a gap in current applications for jurisprudence. The unified design enables scholars to maintain their analytical context while moving between different levels of abstraction. The design can reduce the cognitive load associated with context switching by preserving the semantic relationships between legal materials across visualizations. This integration also facilitates the progressive refinement of legal arguments, as insights developed in detailed analysis can immediately inform and restructure the broader research scope. Our unified design creates a continuous analytical environment that mirrors the non-linear nature of legal reasoning.

Implementing the proposed VA workflow faces challenges.
The suggested visualization techniques for VA may suffer from issues with \emph{visual scalability}. For vast legal datasets, node-link diagrams, Treemaps, and Icicle Plots are of limited use.
However, since the datasets are hierarchically structured, we can progressively unveil parts of the hierarchy to limit the information these visualization techniques need to convey.
Further, leveraging progress in \emph{natural language understanding} remains difficult: legal text data combines domain-specific language~\cite{ruethersRechtstheorieUndJuristische2022} and precise phraseology with intentional ambiguity at times~\cite{ruethersRechtstheorieUndJuristische2022}.
This ambiguity allows for statutory interpretation~\cite{ruethersRechtstheorieUndJuristische2022},  enabling laws to evolve alongside societal norms and values, but is challenging for LLMs.
Although trained on vast datasets, LLMs still struggle to apply these nuances of legal reasoning, particularly open foundation models.
For example, we found that Llama~3.1~70B still sometimes confuses \S{}~212~StGB~(Strafgesetzbuch, i.e., criminal code), which defines homicide for murder~(\S{}~211~StGB).
However, by definition of the criminal code, they are two strictly separate criminal offenses.
While such models can describe the differences in great detail when explicitly asked, during reasoning, they can tend to collapse any distinctions without further warning.
Equally, hallucinations, particularly during the subsumption phase, are critical as LLMs struggle with completeness.
Therefore, it is crucial to perform extensive LLM fine-tuning to align with the specific language of the legal domain.
This also presents ample opportunity for VA design to facilitate collaboration between human scholars and LLMs during the legal reasoning process.

\subsection{Implications for Visual Analytics Design}

The lingering challenges in data navigation, knowledge externalization, and analytical reasoning cannot be sufficiently addressed with AI and computational solutions alone.
Therefore, we advocate for a human-\textit{is}-the-loop design~\cite{endertHumanLoopNew2014}, which positions the VA system as an integral part of the scholar's workflow. This view contrasts designs that rely on the common human-\underline{in}-the-loop~\cite{keimVisualAnalyticsDefinition2008, sachaKnowledgeGenerationModel2014} approach.
While the latter assumes the domain expert to become a part of their VA system, the former prerequisites the VA system to become part of the domain expert and their workflow instead.
To achieve this, the VA design process requires a user-centered approach that honors the workflows of the domain expert, supporting them throughout the sense-making loop~\cite{pirolli2005sensemaking, endertHumanLoopNew2014}.
Therefore, we highlight the following key techniques and higher-level considerations regarding jurisprudence and generalization to other domains.

\textbf{User-Centered Design} --- The development of effective VA design requires adherence to user-centered design principles~\cite{gibbsBuildingBetterDigital2012, hinrichsSpeculativePracticesUtilizing2016}.
These principles promote broader adoption of VA designs that meet specific audience needs.
These needs vary across different application domains, valuing various visual qualities, such as page layout in law.
While these aspects may not seem directly related to the effectiveness or efficiency of VA design, they can significantly influence data navigation in information spaces.
This variation in user requirements presents ample opportunity for employing design studies or conducting pilot interviews with domain experts.
These methodological approaches allow for gaining critical insights into the audience's specific needs, workflows, and mental models, like the importance of tacit knowledge, the lack of contextualized search, and the three-phase research process, which we discussed in~\autoref{section:interviews}.

\textbf{Knowledge Externalization} --- In application domains, subject matter experts possess tacit knowledge regarding domain-specific problems.
It encompasses implicit understanding, experiential insights, and intuitive reasoning processes not readily encoded in explicit knowledge.
The externalization process captures these domain-specific analytical approaches in shareable analytical artifacts.
If not articulated, VA design inherently lacks access to this knowledge, potentially compromising analytical effectiveness.
While explicit knowledge may be available in substantial quantities through text documents, for example, integrating with tacit knowledge is imperative to enhance analytical outcomes~\cite{federicoRoleExplicitKnowledge2017}.
VA design must strive to externalize tacit knowledge through user interaction, KGs, and analytical artifacts.

\textbf{Provenance Tracking} --- Analytical artifacts manifest knowledge externalization processes within VA systems.
These artifacts serve as persistent representations of analytical reasoning and decision-making trajectories throughout the analytical workflow~\cite{perez-messinaPersistentInteractionUserGenerated2024}.
Since users often struggle to maintain holistic awareness of past analytical steps, access to provenance information is essential for branching and revising the analysis process.
VA design must facilitate the creation and storage of analytical artifacts through comprehensive provenance tracking, fostering the user's awareness of the analysis process.
Concurrently, these artifacts can be systematically utilized to guide subsequent analytical processes.
VA design must exploit this recursive employment of analytical artifacts to create a self-reinforcing system that progressively refines the analytical workflow based on accumulated insights and knowledge.

\textbf{Natural Language Inference} --- While mining user interactions with VA interfaces can constitute guidance, it presents inherent methodological challenges in deducing higher-order analytical objectives~\cite{gotzCharacterizingUsersVisual2008}.
Interaction patterns alone provide insufficient semantic context for accurately inferring analytical goals that guide user behavior.
With recent advancements in LLMs, these models demonstrate sophisticated semantic parsing and intention recognition capabilities that can be leveraged to extract structured analytical objectives from natural language~\cite{zhaoLEVAUsingLarge2025}.
VA design must enable users to articulate these analytical objectives through natural language interfaces to capture nuanced analytical intentions while going beyond a simple chat-based interface design.

\textbf{Unified Design} --- To achieve effectiveness and efficiency in addressing domain-specific problems, VA design must integrate the aforementioned key techniques.
User-centered design facilitates identifying and formalizing domain-specific workflows, as demonstrated through our structured interviews with subject matter experts~(see \autoref{section:interviews}).
Such approaches reveal critical insights into domain-specific analytical processes that may not be apparent through conventional design approaches.
Likewise, knowledge externalization provides access to tacit knowledge, informing VA. 
Provenance tracking can effectively reflect domain-specific workflows by capturing the structure of analytical processes.
Minimizing cognitive load by context switching and unifying analytical functionalities within a coherent interface allows one to optimize cognitive resource utilization and analytical performance.

\subsection{Limitations}
Our unified design approaches legal tech from the VA perspective, adopting a definition of visualization~\cite{cardReadingsInformationVisualization2007} from the visualization community in our literature analysis. In contrast, the legal community uses a broader definition~\cite{khalilQuoVadisVisuelle2014}, leading to papers that visualize legal data but are not considered VA. While not our primary focus, we integrate such visualization concepts—familiar to legal scholars—into our VA design~(see \autoref{section:visual-analytics-for-jurisprudence}).
For example, we employ \textit{Concept Maps}, commonly used to represent cases and causal relations between norms~\cite{cantatoreMakingConnectionsIncorporating2016}, to capture tacit knowledge~(see \autoref{section:va_knowledge_representation}) during legal reasoning. Thus, visualization concepts from law can enrich VA design and foster interdisciplinary research.

\textbf{Tasks} --- To enhance our understanding of how legal scholars operate, we describe different levels of tasks~(see \autoref{section:tasks}). These levels are supported by research on applications in legal tech, often emerging from the intersection of law and computer science. Consequently, the labels for these task levels resemble technical terms from VA, such as information retrieval. However, these tasks are grounded in the operations of scholars as described in literature from the legal domain~(see \autoref{section:how-legal-scholars-operate}). Additionally, these tasks are linked to the three-phase workflow~(see \autoref{section:workflow}), which is based on interviews with legal scholars~(see \autoref{section:interviews}).

\textbf{Expert Interviews} --- These results of the subject matter expert interviews~(see \autoref{section:interviews}) take common workflows and recurring challenges faced by German scholars from various legal disciplines into account.
Still, they are not exhaustive, as further disciplines exist in the German legal system and beyond.
To accommodate the differences between legal disciplines, we have decided on a semi-structured interview, where we adapt the follow-up questions to the interviewees' backgrounds and experiences.
Although we assume differences in the details, we expect the significant challenges distilled in \autoref{section:visual-analytics-for-jurisprudence} to be similar, as they concern all disciplines.
We elicit our observations from subject matter experts who work exclusively with the German civil law system and European legislation.
So far, we have not incorporated experiences from countries that practice civil law or other legal systems like common law~(see \autoref{section:foundations-and-background}).

\textbf{Case Study} --- To illustrate the three-phase workflow design~(see \autoref{section:workflow}) and to critically discuss traditional UI for legal research in our case study~(see \autoref{section:visual-analytics-for-jurisprudence}), we have focused on the \emph{Bundespolizeigesetz} (Federal Police Act).
This specific legal research context highlights the challenges described in a broader context by participants during the interviews~(see \autoref{section:interviews}) using an example.
Challenges arise from accessibility issues, navigational complexity, the inefficiency of the search interface, and the reliance on tacit knowledge, which extend to other laws and research contexts.

\textbf{Generalization} --- This generalization allows our case study to transcend the specific example chosen and the German legal system.
To illustrate the traditional UI used in jurisprudence as part of our case study~(see \autoref{section:visual-analytics-for-jurisprudence}), we use \beckOnline{} as an example. Although \beckOnline{} is just one of many commercial and scientific applications for jurisprudence, it virtually has a monopoly on legal research by practitioners. Additionally, it features a design similar to applications such as \textit{Westlaw}~\cite{thomsonreutersWestlawClassic} and \textit{PACER}~\cite{administrativeofficeoftheu.s.courtsPACERPublicAccess}, which helps generalize our findings to case law.
Further, our work lacks empirical, quantitative evaluations of the proposed VA workflows as a prototype, as empirical evaluations from other domains might not generalize to the legal field.

\subsection{Future Work and Generalization}
\label{sec:generalization}

To address the limited evaluation, we plan to implement the proposed VA workflow for a systematic evaluation.
In this context, corresponding user studies should involve controlled experiments comparing the efficacy and usability.
Additionally, the experiments should assess the cognitive load experienced when utilizing interactive visualizations.

Beyond legal scholars’ evaluations, future research should focus on communicating law and legal concepts to the general public.
This necessitates research into adaptive, personalized visualization techniques that cater to the user’s expertise level rather than assuming the expertise of legal scholars.
Personalized visualization also allows us to explore the adoption of specific visual metaphors prevalent in legal sub-disciplines like administrative law or criminal law.

Our work is scoped to the German civil law system, with qualitative insights derived from nine legal experts representing various legal disciplines within this system.
Consequently, the proposed VA design fits this context but offers a methodological framework that could be replicated in other legal environments within the civil law system.
While these environments work similarly, countries sometimes have substantial differences regarding style, structure, and methodology.
For example, EU legislation and German law differ in articulating their goals.
In EU legislation, goals are clearly outlined, while German law lacks such explicit statements, sometimes to be found in the law justifications.
Additionally, the two systems vary in their teleological and systematic interpretations.
There are also differences in the integration of certain aspects of case law into EU legislation as a supranational legal system.
Future work should explore the particularities of extending our approach to other legislative systems and, in particular, to case law, with considerable differences.
For replication, researchers would need to interview domain experts familiar with the legal workflows in that system to validate the adaptation of our three-phase workflow.
Additionally, the researchers would need to have visualization expertise to verify that the adaption still aligns with the proposed VA design.

While we have presented the textual analysis with a strong focus on jurisprudence, some methods and solutions exhibit substantial overlap with workflows in related, text-based research domains~\cite{zengVIStoryInteractiveStoryboard2021, tianLitVisVisualAnalytics2023}.
These workflows share core elements of data navigation, knowledge representation, and analytical reasoning, as outlined in \autoref{section:visual-analytics-for-jurisprudence}.
Consequently, our VA workflow (see \autoref{section:addressing-jurisprudence}) can be transferred onto these usage scenarios, albeit with domain-specific adaptations.
We suppose that the workflow phases and the visualization techniques used to support data navigation, knowledge representation, and analytical reasoning are domain-agnostic, while the underlying data, its hierarchical structure, and details of the reasoning processes are domain-specific.
Hence, similar to jurisprudence, the domain transferred to must offer accessible data sets that impose a hierarchical structure on the domain semantics.

Overall, our VA workflow can provide insights and a starting ground for visualization researchers in domains where tacit knowledge remains a driving force for analytical reasoning and decision-making.

\section{Conclusion}

In this work, we explore the integration of Visual Analytics (VA) into jurisprudence, addressing the unique challenges of legal texts' complex, formal, and hierarchical nature.
Through semi-structured interviews with legal experts, we identify a typical workflow in jurisprudence and analyze its commonalities with workflows in other disciplines.
This analysis reveals an abstraction in the sense-making loop, emphasizing the need for VA systems that integrate data navigation, knowledge representation, and analytical reasoning within a unified design.
To address these gaps, we propose a human-is-the-loop VA design that leverages legal scholars' tacit knowledge, enhancing their ability to navigate, interpret, and reason with legal documents.
We critically discuss the necessary transformations from a traditional user interface commonly employed to a VA-based interface, alongside its challenges.
Our proposed approach facilitates informed legal reasoning by combining human expertise with machine intelligence.
It provides a scalable framework for other text-intensive domains and the blueprint for future development of such systems.
Our findings extend beyond jurisprudence, offering valuable insights into the design of VA systems for related fields.
This work lays a foundation for future advancements in knowledge-assisted VA systems by addressing the interplay between tacit knowledge and sense-making processes.


\begin{backmatter}



\section*{Author Contributions}
The manuscript was coordinated by Daniel Fürst, with important contributions from Maximilian T. Fischer on all sections. Mennatallah El-Assady and Daniel A. Keim contributed with their expertise in the domain, providing useful feedback during the process. All authors reviewed the manuscript.

\section*{Funding}
This work has been funded by the Federal Ministry of Education and Research (BMBF) in VIKING (13N16242), the Deutsche Forschungsgemeinschaft (DFG, German Research Foundation) in RATIO-CUEPAQ (455910360), and under Germany’s Excellence Strategy - EXC 2117 - 422037984.

\section*{Declarations}

\subsection*{Conflict of interest}
There is no conflict of interest or competing interests from any of the authors.

\subsection*{Consent to participate}
All participants were informed and gave their consent to participate in the study. Participants were informed about the use of the questionnaire/interview data for an article.

\subsection*{Consent for publication}
All participants gave their consent to publish the data collected in the surveys and related to their participation in the study. We do not publish any identifying information about the participants.

\vspace*{2em}

\printaddresses

\printbibliography
\end{backmatter}

\end{document}